\documentclass[10pt,twocolumn,letterpaper]{article}

\usepackage{cvpr}
\usepackage{times}
\usepackage{epsfig}
\usepackage{graphicx}
\usepackage{amsmath}
\usepackage{amssymb}
\usepackage[numbers,sort]{natbib}

\usepackage{support-caption}
\usepackage{comment, array, multirow, pifont, bm, mathtools, subcaption, float, graphicx, xcolor, placeins, booktabs, xfrac, titling, scalerel, enumerate, enumitem}
\newcolumntype{M}[1]{>{\centering\arraybackslash}m{#1}}
\newcommand{\xmark}{\ding{55}}
\DeclarePairedDelimiter\autobracket{(}{)}
\newcommand{\br}[1]{\autobracket*{#1}}
\definecolor{green(ryb)}{rgb}{0.4, 0.69, 0.2}
\definecolor{ao(english)}{rgb}{0.0, 0.5, 0.0}
\definecolor{cornellred}{rgb}{0.7, 0.11, 0.11}

\usepackage[pagebackref=true,
breaklinks=true,
colorlinks,
bookmarks=false,
citecolor = ao(english),
linkcolor = cornellred,
]{hyperref}

\cvprfinalcopy 

\def\cvprPaperID{9758} 

\begin{document}

\title{Non-Adversarial Video Synthesis with Learned Priors}

\author{Abhishek Aich$^{\dagger,*}$, Akash Gupta$^{\dagger, }$\thanks {Joint first authors}~, Rameswar Panda$^{\ddagger}$, Rakib Hyder$^{\dagger}$,\\
M. Salman Asif$^{\dagger}$, Amit K. Roy-Chowdhury$^{\dagger}$\\
University of California, Riverside$^{\dagger}$, IBM Research AI, Cambridge$^{\ddagger}$\\
{\tt \small \{aaich001@, agupt013@, rpand002@, rhyde001@, sasif@ece.,  amitrc@ece.\}ucr.edu}
}
\date{\vspace{-3ex}}
\maketitle

\begin{abstract}
Most of the existing works in video synthesis focus on generating videos using adversarial learning. Despite their success, these methods often require input reference frame or fail to generate diverse videos from the given data distribution, with little to no uniformity in the quality of videos that can be generated. Different from these methods, we focus on the problem of generating videos from latent noise vectors, without any reference input frames. To this end, we develop a novel approach that jointly optimizes the input latent space, the weights of a recurrent neural network and a generator through non-adversarial learning. 
Optimizing for the input latent space along with the network weights allows us to generate videos in a controlled environment, i.e., we can faithfully generate all videos the model has seen during the learning process as well as new unseen videos. Extensive experiments on three challenging and diverse datasets well demonstrate that our proposed approach generates superior quality videos compared to the existing state-of-the-art methods.
\end{abstract}

\section{Introduction}\label{sec:intro}
Video synthesis is an open and challenging problem in computer vision. As literature suggests, a deeper understanding of spatio-temporal behavior of video frame sequences can directly provide insights in choosing priors, future prediction, and feature learning \cite{vondrick2016generating, p2pvg2019}. Much progress has been made in developing variety of ways to generate videos which can be classified into broadly two categories: class of video generation methods which require random latent vectors without any reference input pixel \cite{vondrick2016generating, saito2017temporal, tulyakov2018mocogan}, and class of video generation methods which do depend on reference input pixels \cite{p2pvg2019, he2018probabilistic, siarohin2019animating}. Current literature contains methods mostly from the second class which often requires some human intervention \cite{p2pvg2019, he2018probabilistic}.

\begin{figure}
    \centering
    \includegraphics[width=0.47\textwidth]{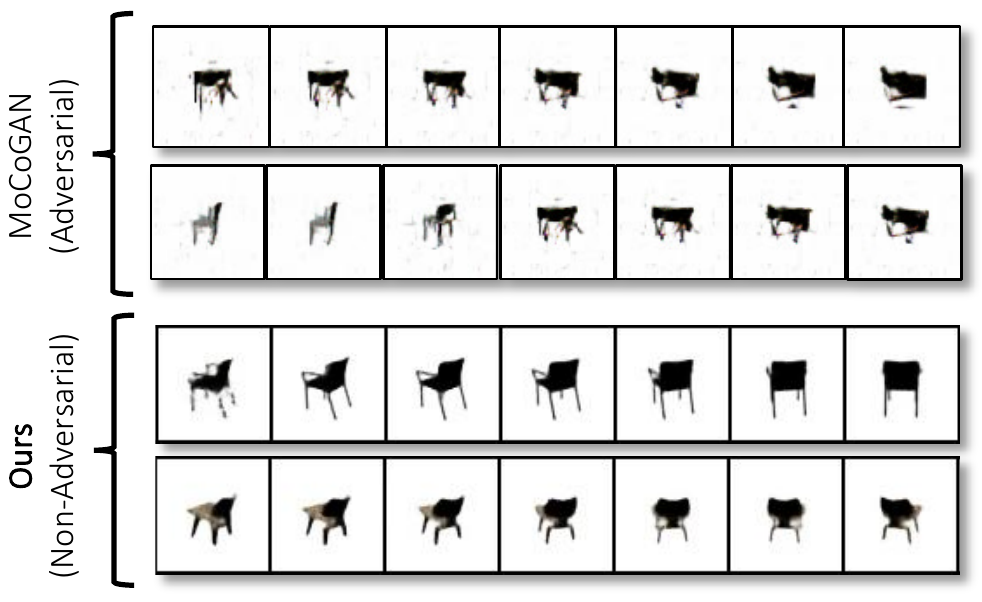}
    \caption{\textbf{Comparison of proposed non-adversarial approach to one representative adversarial approach (MoCoGAN~\cite{tulyakov2018mocogan}) on the Chair-CAD~\cite{aubry2014seeing} dataset.} \textit{Top}: MoCoGAN often generates blurry frames including similar type of chairs for different videos as the time step increases. \textit{Bottom}: Our approach\protect\footnotemark, on the other hand, generates relatively sharper frames, maintaining consistency with the type of chairs unique to each video in the dataset.} 
    \label{fig:first_fig}
     \vspace{-1em}
\end{figure} 
\footnotetext{Project page: \textcolor{blue}{https://abhishekaich27.github.io/navsynth.html}}
In general, Generative Adversarial Networks (GANs)~\cite{goodfellow2014generative}
 have shown remarkable success in various kinds of video modality problems \cite{liang2017dual, kwon2019predicting, lotter2016deep, saito2017temporal}. Initially, video generation frameworks predominantly used GANs to synthesize videos from latent noise vectors. For example, VGAN \cite{vondrick2016generating} and TGAN \cite{saito2017temporal} proposed  generative models that synthesize videos from random latent vectors with deep convolutional GAN. Recently, MoCoGAN \cite{tulyakov2018mocogan} proposed to decompose a video into content and motion parts using a generator guided by two discriminators. During testing, these frameworks generate videos that are captured in the range of the trained generator, by taking random latent vectors.  While all these methods have obtained reasonable performance on commonly used benchmark datasets, they utilize adversarial learning to train their models and hence, inherit the shortcomings of GANs. Specifically, GANs are often very sensitive to multiple factors such as random network initialization, and type of layers employed to build the network \cite{li2017towards, salimans2016improved}. Some infamous drawbacks of GANs are mode-collapse (i.e., able to generate only some parts of the data distribution: see Fig.~\ref{fig:first_fig} for an example) and/or vanishing generator gradients due to discriminator being way better in distinguishing fake samples and real samples \cite{arjovsky2017wasserstein}.

\begin{figure*}[t]
\centering
\includegraphics[width=0.87\textwidth]{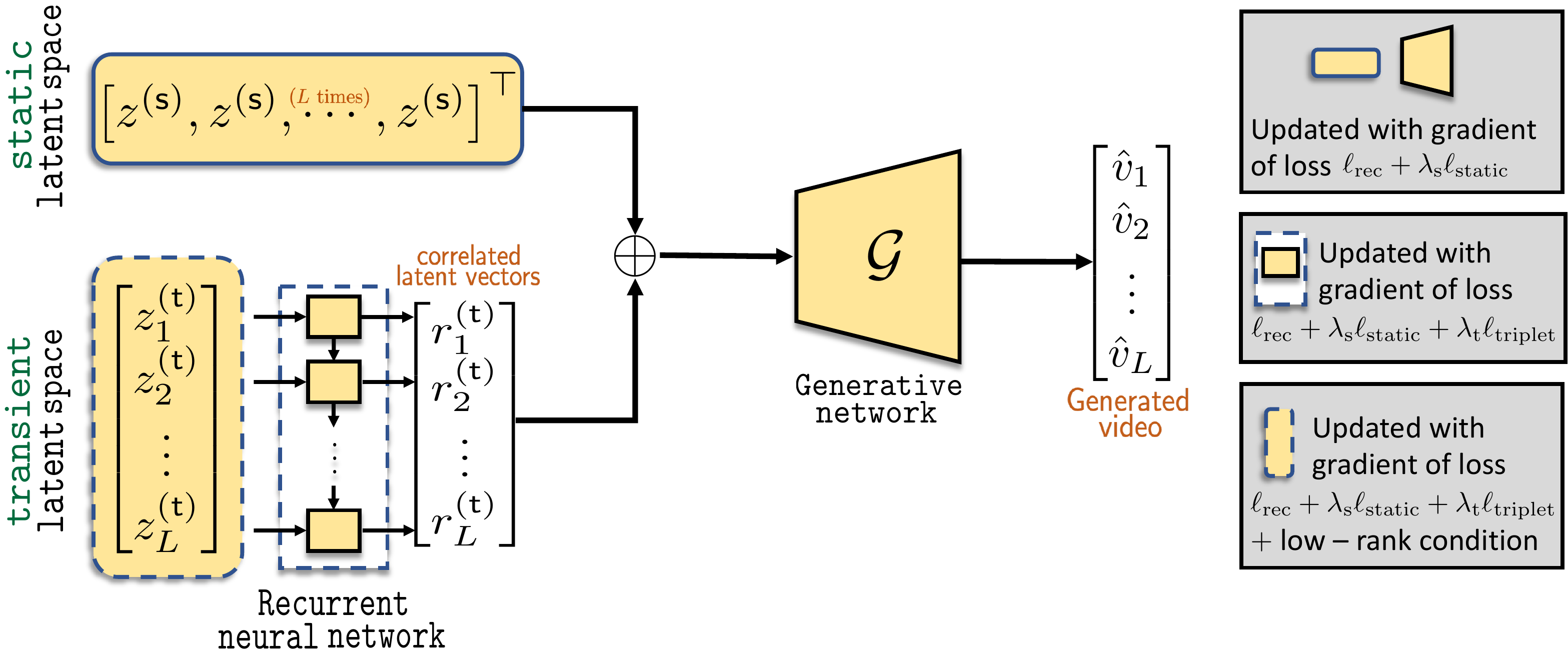}
\caption{\textbf{Overview of the proposed method.} Videos can be broken down into two main parts: static and transient components. To capture this, we map a video (with \textit{L} frame sequence) into two learnable latent spaces. We jointly learn the static latent space and the transient latent space along with the network weights. We then use these learned latent spaces to generate videos at the inference time. See Sec.~\ref{sec:prop_method} for more details.}
\label{fig:intro}
\vspace{-1em}
\end{figure*}

Non-adversarial approaches \cite{bojanowski2018optimizing, li2018implicit, hoshen2019non} have recently been explored to tackle these challenges. For example, Generative Latent Optimization (GLO) \cite{bojanowski2018optimizing} and Generative Latent Nearest Neighbor (GLANN) \cite{hoshen2019non} investigate the importance of inductive bias in convolutional networks by disconnecting the discriminator for a non-adversarial learning protocol of GANs. These works show that without a discriminator, a generator can be learned to map the training images in the given data distribution to a lower dimensional latent space that is learned in conjunction with the weights of the generative network. Such procedure not only avoids the mode-collapse problem of the generators, but also provides the user an optimized low dimensional latent representation (embedding) of the data in contrast with the random latent space as in GANs. Recently Video-VAE \cite{he2018probabilistic} proposed to use Variational Auto-Encoder (VAE) for conditional video synthesis, either by randomly generating or providing the first frame to the model for synthesizing a video. However, the quality of generated videos using Video-VAE often depends on the provided input frame. Non-adversarial video synthesis without any visual inputs still remains as a novel and rarely addressed problem.

\emph{In this paper}, we propose a novel non-adversarial framework to generate videos in a controllable manner without any reference frame. Specifically, we propose to synthesize videos from two optimized latent spaces, one providing control over the static portion of the video (\textit{static latent space}) and the other over the transient portion of the video (\textit{transient latent space}). We propose to jointly optimize these two spaces while optimizing the network (a generative and a recurrent network) weights with the help of regression-based reconstruction loss and a triplet loss. 

Our approach works as follows. During training, we jointly optimize over network weights and latent spaces (both static and transient) and obtain a common transient latent space, and individual static latent space dictionary for all videos sharing the same class (see Fig.~\ref{fig:intro}). During testing, we randomly choose a static vector from the dictionary, concatenate it with the transient latent vector and generate a video. This enables us to obtain a controlled environment of diverse video generation from learned latent vectors for each video in the given dataset, while maintaining almost uniform quality. In addition, the proposed approach also allows a concise video data representation in form of learned vectors, frame interpolation (using a low rank constraint introduced in \cite{hyder2020low}), and generation of videos unseen during the learning paradigm. 

The key contributions of our work are as follows.
\begin{itemize}[leftmargin=*]
\item We propose a novel framework for generating a wide range of diverse videos from learned latent vectors without any conditional input reference frame with almost uniform visual quality. Our framework obtains a latent space dictionary on both static and transient portions for the training video dataset, which enables us to generate even unseen videos with almost equal quality by providing combinations of static and transient latent vectors that were not part of training data. 
\item Our extensive experiments on multiple datasets well demonstrate that the proposed method, without the adversarial training protocol, has better or at par, performance with current state-of-the-art methods \cite{vondrick2016generating, saito2017temporal, tulyakov2018mocogan}. Moreover, we do not need to optimize the (multiple) discriminator networks as in previous methods \cite{vondrick2016generating, saito2017temporal, tulyakov2018mocogan} which offers a computational advantage.
\end{itemize}	 

\section{Related Works}\label{sec: rel_work}
Our work relates to two major research directions:
video synthesis and non-adversarial learning. This section focuses on some representative methods closely related to our work.

\subsection{Video Synthesis} Video synthesis has been studied from multiple perspectives \cite{vondrick2016generating, saito2017temporal, tulyakov2018mocogan, he2018probabilistic, siarohin2019animating} (see Tab.~\ref{tab:compare_methods} for a categorization of existing methods). VGAN \cite{vondrick2016generating} demonstrates that a video can be divided into foreground and background using deep neural networks. TGAN \cite{saito2017temporal} proposes to use a  generator to capture temporal dynamics by generating correlated latent codes for each video frame and then using an image generator to map each of these latent codes to a single frame for the whole video. MoCoGAN \cite{tulyakov2018mocogan} presents a simple approach to separate content and motion latent codes of a video using adversarial learning. The most relevant work for us is Video-VAE \cite{he2018probabilistic} that extends the idea of image generation to video generation using VAE by proposing a structured latent space in conjunction with the VAE architecture for video synthesis. While this method doesn't require a discriminator network, it depends on reference input frame to generate a video. In contrast, our method proposes a efficient framework for synthesizing videos from learnable latent vectors without any input frame. This gives a controlled environment for video synthesis that even enables us to generate visually good quality unseen videos through combining static and transient parts.

\begin{table}   
\begin{center}\small{
\begin{tabular}{M{1.7cm}|M{1.5cm}|M{1.4cm}|M{1.8cm}}
\toprule[1.2pt]
\multicolumn{1}{c|}{\multirow{2}{*}{Methods}} & \multicolumn{3}{c}{Settings}\\
\cline{2-4}
 & \small{Adversarial learning?} & \small{Input frame?} & \small{Input latent vectors?} \\
\midrule
VGAN \cite{vondrick2016generating} & \checkmark & \xmark & \checkmark (random)\\
\hline
TGAN \cite{saito2017temporal} & \checkmark & \xmark & \checkmark (random) \\
\hline
MoCoGAN \cite{tulyakov2018mocogan} & \checkmark & \xmark & \checkmark (random)\\
\hline
Video-VAE \cite{he2018probabilistic} & \xmark & \checkmark & \checkmark (random) \\
\hline
\textbf{Ours} & \xmark & \xmark & \checkmark (learned)\\
\bottomrule[1.2pt]
\end{tabular}}
\end{center}
\caption{\textbf{Categorization of prior works in video synthesis.} Different from existing methods, our model doesn't require a discriminator, or any reference input frame. However, since we have learned latent vectors, we have control of the kind of videos the model should generate.}
\label{tab:compare_methods}
\vspace{-1.1em}
\end{table}

\subsection{Non-Adversarial Learning} Generative adversarial networks, as powerful as they are in pixel space synthesis, are also difficult to train. This is owing to the saddle-point based optimization game between the generator and the discriminator. On top of the challenges discussed in the previous section, GANs require careful user driven configuration tuning which may not guarantee same performance for every run. Some techniques to make the generator agnostic to described problems have been discussed in \cite{salimans2016improved}. The other alternative to the same has given rise to non-adversarial learning of generative networks \cite{bojanowski2018optimizing, hoshen2019non}. Both \cite{bojanowski2018optimizing, hoshen2019non} showed that properties of convolutional GANs can be mimicked using simple reconstruction losses while discarding the discriminator. 
 
While there has been some work on image generation from learned latent vectors \cite{bojanowski2018optimizing, hoshen2019non}, our work significantly differs from these methods as we do not map all the frames pixel-wise in a given video to the same latent distribution. This is because doing so would require a separate latent space (hence, a separate model for each video) for all the videos in a given dataset, and performing any operation in that space would naturally become video specific. Instead, we divide the latent space of videos sharing the same class into two parts - static and transient. This gives us a dictionary of static latent vectors for all videos and a common transient latent subspace. Hence, any random video of the dataset can now be represented by the combination of one static vector (which remains same for all frames) and the common transient subspace.  

\section{Formulation}\label{sec:prop_method} 
Define a video clip $\mathsf{V}$ represented by $L$ frames as
$\mathsf{V} = \begin{bmatrix}v_1, v_2, \cdots, v_L
    \end{bmatrix}$. Corresponding to each frame $v_i$, let there be a point $z_i$ in latent space $\mathcal{Z}_\mathsf{V}\in\mathbb{R}^{D\times L}$ such that
\begin{align}\label{2}
    \mathcal{Z}_{\mathsf{V}} = \begin{bmatrix}z_1, z_2, \cdots, z_L
    \end{bmatrix}
\end{align}
which forms a path of length $L$. We propose to disentangle a video into two parts: a static constituent, which captures the constant portion of the video common for all frames, and a transient constituent which represents the temporal dynamics between all the frames in the video. Hence, let $\mathcal{Z}_\mathsf{V}$ be decomposed as $\mathcal{Z}_\mathsf{V} = [\mathcal{Z}_\mathsf{s}^\top ,\mathcal{Z}_\mathsf{t}^\top]^\top$ where $\mathcal{Z}_\mathsf{s}\in\mathbb{R}^{D_\mathsf{s}\times L}$ represents the static  subspace and $\mathcal{Z}_\mathsf{t}\in\mathbb{R}^{D_\mathsf{t}\times L}$ represents the transient subspace with $D = D_\mathsf{s} + D_\mathsf{t}$. Thus $\{z_i\}_{i=1}^L$ in (\ref{2}) can be expressed as $z_i = \begin{bmatrix}z^{(\mathsf{s})\top}_{i} ,z^{(\mathsf{t})\top}_{i}\end{bmatrix}^\top~\forall~i=1, 2,\cdots,L$. Next assuming that the video is of short length, we can fix  $z^{(\mathsf{s})}_{i} = z^{(\mathsf{s})}$ for all frames after sampling only once. Therefore, (\ref{2}) can be expressed as
\begin{align}\label{3}
    \mathcal{Z}_\mathsf{V} = \begin{bmatrix}\begin{bmatrix}z^{(\mathsf{s})} \\ z^{(\mathsf{t})}_{1}\end{bmatrix}, \begin{bmatrix}z^{(\mathsf{s})} \\ z^{(\mathsf{t})}_{2}\end{bmatrix}, \cdots, \begin{bmatrix}z^{(\mathsf{s})} \\ z^{(\mathsf{t})}_{L}\end{bmatrix}
    \end{bmatrix}
\end{align}

The transient portion will represent the motion of a given video. Intuitively, the latent vectors corresponding to this transient state should be correlated, or in other words, will form a path between $z^{(\mathsf{t})}_{1}$ and $z^{(\mathsf{t})}_{L}$. 
Specifically, the frames in a video are correlated in time and hence a frame $v_i$ at time $i = T$ is a function of all previous frames $\{v_i\}_{i=1}^{T-1}$. As a result, their corresponding transient representation should also exhibit such a trajectory. This kind of representation of latent vectors can be obtained by employing a Recurrent Neural Network (RNN) where output of each cell of the network is a function of its previous state or input. Denote the RNN as $\mathcal{R}$ with weights $\theta$. Then, the RNN output $\mathcal{R}\br{z_i} = \{r^{(\mathsf{t})}_{i}\}~\forall~i = 1, 2, \cdots, L$ is a sequence of correlated variables representing the transient state of the video.

\subsection{Learning Network Weights}
Define a generative network $\mathcal{G}$ with weights represented by $\bm{\gamma}$.  $\mathcal{G}$ takes latent vectors sampled from $\mathcal{Z}_\mathsf{V}$ as input and predicts up to $L$ frames of the video clip. For a set of $N$ videos, initialize set of \textit{D}-dimensional vectors $\mathcal{Z}_\mathsf{V}$ to form the pair \Big{\{}\big{(}$\mathcal{Z}_{\mathsf{V}_1}, \mathsf{V}_1$\big{)}, \big{(}$\mathcal{Z}_{\mathsf{V}_2}, \mathsf{V}_2$\big{)}, $\cdots$, \big{(}$\mathcal{Z}_{\mathsf{V}_N}, \mathsf{V}_N$\big{)} \Big{\}}. More specifically from (\ref{3}), defining $\textbf{z}^{(\mathsf{s})} = \begin{bmatrix}z^{(\mathsf{s})}, z^{(\mathsf{s})}, \cdots, z^{(\mathsf{s})}\end{bmatrix}\in\mathbb{R}^{D_\mathsf{s}\times L}$, and $\textbf{z}^{(\mathsf{t})} = \begin{bmatrix}z_1^{(\mathsf{t})}, z_2^{(\mathsf{t})}, \cdots, z_L^{(\mathsf{t})}\end{bmatrix}\in\mathbb{R}^{D_\mathsf{t}\times L}$, we will have the pairs
\begin{align*}
\Bigg{\{}\br{\begin{bmatrix}\textbf{z}^{(\mathsf{s})} \\ \textbf{z}^{(\mathsf{t})}\end{bmatrix}_{1}, \mathsf{V}_1}, \br{\begin{bmatrix}\textbf{z}^{(\mathsf{s})} \\ \textbf{z}^{(\mathsf{t})}\end{bmatrix}_{2}, \mathsf{V}_2}, \cdots, \br{\begin{bmatrix}\textbf{z}^{(\mathsf{s})} \\ \textbf{z}^{(\mathsf{t})}\end{bmatrix}_{N}, \mathsf{V}_N} \Bigg{\}}.
\end{align*} 
With these pairs, we propose to optimize the weights $\gamma$, $\theta$, and input latent vectors $\mathcal{Z}_\mathsf{V}$ (sampled once in the beginning of training) in the following manner. For each video $\mathsf{V}_j$, we jointly optimize for $\theta, \gamma$, and $\{\mathcal{Z}_{\mathsf{V}_j}\}_{j=1}^N$ for every epoch in two stages:
\begin{subequations}\label{4}
        \renewcommand{\theequation}{\theparentequation.\arabic{equation}}
\begin{alignat}{3}
& \text{Stage 1}:   \quad && \min_{\gamma}~\ell\br{ \mathsf{V}_j, \mathcal{G}(\mathcal{Z}_{\mathsf{V}_j})\vert\br{\mathcal{Z}_{\mathsf{V}_j},\theta}}\label{4.1}\\
& \text{Stage 2}:   \quad &&  \quad \min_{\mathcal{Z}_\mathsf{V}, \theta}~\ell\br{ \mathsf{V}_j, \mathcal{G}(\mathcal{Z}_{\mathsf{V}_j})\vert\gamma}\label{4.2}
\end{alignat}
\end{subequations}
$\ell(\cdot)$ can be any regression-based loss. For rest of the paper, we will refer to both (\ref{4.1}) and (\ref{4.2}) together as $\min\limits_{\mathcal{Z}_\mathsf{V}, \theta, \gamma}\ell_{\text{rec}}$.

\noindent \textbf{Regularized loss function to capture static subspace.}
The transient subspace, along with the RNN, handles the temporal dynamics of the video clip. To equally capture the static portion of the video, we randomly choose a frame from the video and ask the generator to compare its corresponding generated frame during training. For this, we update the above loss as follows. 
\begin{align}\label{5}
    \min_{\mathcal{Z}_\mathsf{V}, \theta, \gamma}~\br{\ell_{\text{rec}} + \lambda_{\text{s}}\ell_{\text{static}}}
\end{align} 
where $\ell_{\text{static}} = \ell\br{\hat{v}_k, v_k}$ with $\textit{k}\in\{1, 2, \cdots, L\}$, $v_k$ is the ground truth frame, $\hat{v}_k = \mathcal{G}(\textbf{z}_k)$, and $\lambda_{\text{s}}$ is the regularization constant. $\ell_{\text{static}}$ can also be understood to essentially handle the role of image discriminator in \cite{tulyakov2018mocogan, wang2018video} that ensures that the generated frame is close to the ground truth frame. 

\subsection{Learning Latent Spaces}\label{sec:Latent_space}
Non-adversarial learning involves joint optimization of network weights as well as the corresponding input latent space. Apart from the gradients with respect to loss in (\ref{5}), we propose to further optimize the latent space with gradient of a loss based on the triplet condition as follows.
\subsubsection{The Triplet Condition}\label{sec:trip_cond}
\begin{figure}[ht]
    \vspace{-1em} 
    \centering
    \includegraphics[width=0.47\textwidth]{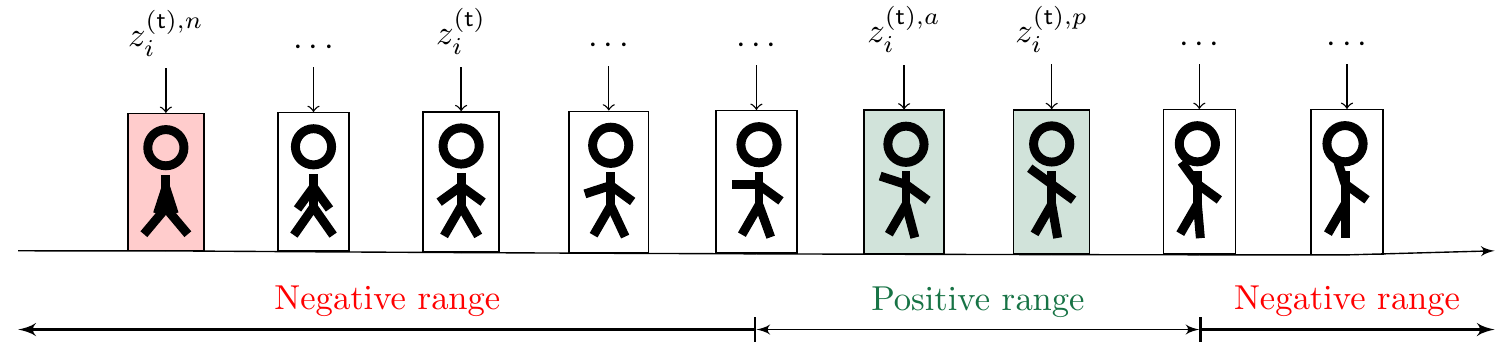}
    \caption{\textbf{Triplet Condition in the transient latent space.} Latent code representation of different frames of short video clips may lie very near to each other in the transient subspace. Using the proposed triplet condition, our model learns to explain the dynamics of similar looking frames and simultaneously map them to distinct latent vectors.}
    \label{fig:triplet_condl}
\end{figure}
Short video clips often have indistinguishable dynamics in consecutive frames which can force the latent code representations to lie very near to each other in the transient subspace. However, an ideal transient space should ensure that the latent vector representation of a frame should only be close to a similar frame than a dissimilar one \cite{schroff2015facenet, sermanet2018time}. To this end, we introduce a triplet loss to (\ref{5}) that ensures that a pair of co-occurring frames $v^a_i$ (anchor) and $v^p_i$ (positive) are closer but distinct to each other in embedding space than any other frame $v^n_i$ (negative) (see Fig.~\ref{fig:triplet_condl}). In this work, positive frames are randomly sampled within a margin range $\alpha$ of the anchor and negatives are chosen outside of this margin range. Defining a triplet set with transient latent code vectors \{$z^{(\mathsf{t}),a}_i, z^{(\mathsf{t}),p}_i, z^{(\mathsf{t}),n}_i$\}, we aim to learn the transient embedding space $\textbf{z}^{(\mathsf{t})}$ such that
\begin{align*}
    \Vert z^{(\mathsf{t}),a}_i - z^{(\mathsf{t}),p}_i\Vert_2^2 + \alpha < \Vert z^{(\mathsf{t}),a}_i - z^{(\mathsf{t}),n}_i\Vert_2^2
\end{align*}
$\forall~\{z^{(\mathsf{t}),a}_i, z^{(\mathsf{t}),p}_i, z^{(\mathsf{t}),n}_i\}\in\Gamma$, where $\Gamma$ is the set of all possible triplets in $\textbf{z}^{(\mathsf{t})}$. With the above regularization, the loss in (\ref{5}) can be written as
\begin{align}\label{6}
    &\qquad\qquad\qquad\min_{\mathcal{Z}_\mathsf{V}, \theta, \gamma}~\br{\ell_{\text{rec}} + \lambda_{\text{s}}\ell_{\text{static}}}\nonumber\\
    &\text{s.t.}\quad \Vert z^{(\mathsf{t}),a}_i - z^{(\mathsf{t}),p}_i\Vert_2^2 + \alpha < \Vert z^{(\mathsf{t}),a}_i - z^{(\mathsf{t}),n}_i\Vert_2^2
\end{align} 
where $\alpha$ is a hyperparameter that controls the margin while selecting positives and negatives.

\subsection{Full Objective Function}
For any choice of differentiable generator $\mathcal{G}$, the objective (\ref{5}) will be differentiable with respect to $\mathcal{Z}_\mathsf{V}$, and $\br{\gamma, \theta}$ \cite{bora2017compressed}. We initialize $\mathcal{Z}_\mathsf{V}$ by sampling them from two different Gaussian distributions for both static and transient latent vectors. We also ensure that the latent vectors $\mathcal{Z}_\mathsf{V}$ lie on the unit $\ell_2$ sphere, and hence, we project $\mathcal{Z}_\mathsf{V}$ after each update by dividing its value by $\mathsf{max}\br{1, \Vert\mathcal{Z}_\mathsf{V}\Vert}$ \cite{bojanowski2018optimizing}, where $\mathsf{max}\br{\cdot}$ returns maximum among the set of given elements. Finally, the complete objective function can be written as follows.
\begin{align}\label{6i}
    &\min_{\mathcal{Z}_\mathsf{V}, \theta, \gamma}~\br{\ell_{\text{rec}} + \lambda_{\text{s}}\ell_{\text{static}} + \lambda_{\text{t}}\ell_{\text{triplet}}}
\end{align}
where $\ell_{\text{static}} = \ell\br{\hat{v}_k, v_k}$, $\lambda_{\text{t}}$ is a regularization constant for the triplet loss, and $\ell_{\text{triplet}} = \mathsf{max}\br{\Vert z^{(\mathsf{t}),a}_i - z^{(\mathsf{t}),p}_i\Vert_2^2 + \alpha - \Vert z^{(\mathsf{t}),a}_i - z^{(\mathsf{t}),n}_i\Vert_2^2, 0}$. The weights of the generator $\gamma$ and static latent vector $z^{(\mathsf{s})}$ are updated by gradients of the losses $\ell_{\text{rec}}$ and  $\ell_{\text{static}}$. The weights. $\theta$, of the RNN, and transient latent vectors $\textbf{z}^{(\mathsf{t})}$ are updated by gradients of the losses $\ell_{\text{rec}}$, $\ell_{\text{static}}$ and $\ell_{\text{triplet}}$.
\vspace{-0.25cm}
\subsubsection{Low Rank Representation for Interpolation}
The objective of video frame interpolation is to synthesize non-existent frames in-between the reference frames. While the triplet condition ensures that similar frames have their transient latent vectors nearby, it doesn't ensure that they lie on a manifold where simple linear interpolation will yield latent vectors that generate frames with plausible motion compared to preceding and succeeding frames \cite{hyder2020low,bojanowski2018optimizing}. This means that the transient latent subspace can be represented in a much lower dimensional space compared to its larger ambient space. So, to enforce such a property, we project the latent vectors into a low dimensional space while learning them along with the network weights, first proposed in \cite{hyder2020low}. Mathematically, the loss in (\ref{6i}) can be written as
\begin{align}\label{7}
    &\min_{\mathcal{Z}_\mathsf{V}, \theta, \gamma}~\br{\ell_{\text{rec}} + \lambda_{\text{s}}\ell_{\text{static}} + \lambda_{\text{t}}\ell_{\text{triplet}}}\nonumber\\
    &~~\text{s.t.}\qquad~~~~\mathsf{rank}\br{\textbf{z}^{(\mathsf{t})}} = \rho
\end{align} 
where $\mathsf{rank}\br{\cdot}$ indicates rank of the matrix and $\rho$ is a hyper-parameter that decides what manifold $\textbf{z}^{(\mathsf{t})}$ is to be projected on. 
We achieve this by reconstructing $\textbf{z}^{(\mathsf{t})}$ matrix from its top $\rho$ singular vectors in each iteration \cite{friedman2001elements}.
Note that, we only employ this condition for optimizing the latent space for the frame interpolation experiments in Sec.~\ref{sec:Frame_inter}. 

\section{Experiments}
In this section, we present extensive experiments to demonstrate the effectiveness of our proposed approach in generating videos through learned latent spaces.

\subsection{Datasets}\label{sec:dataset}

We evaluate the performance of our approach using three publicly available datasets which have been used in many prior works \cite{he2018probabilistic, vondrick2016generating, tulyakov2018mocogan}.

\textbf{Chair-CAD }\cite{aubry2014seeing}.  
This dataset consists of total 1393 chair-CAD models, out of which we randomly choose 820 chairs for our experiments with the first 16 frames, similar to \cite{he2018probabilistic}. The rendered frame in each video for all the models are center-cropped and then resized to $64 \times 64 \times 3$ pixels. We obtain the transient latent vectors for all the chair models with one static latent vectors for the training set.

\textbf{Weizmann Human Action }\cite{gorelick2007actions}. This dataset provides 10 different actions performed by 9 people, amounting to 90 videos. Similar to Chair-CAD, we center-cropped each frame, and then resized to $64 \times 64 \times 3$ pixels. For this dataset, we train our model to obtain nine static latent vectors (for nine different identities) and ten transient latent vectors (for ten different actions) for videos with 16 frames each.   

\textbf{Golf scene dataset }\cite{vondrick2016generating}. Golf scene dataset \cite{vondrick2016generating} contains 20,268 golf videos with $128 \times 128 \times 3$ pixels which further has 583,508 short video clips in total. We randomly chose 500 videos with 16 frames each and resized the frames to $64 \times 64 \times 3$ pixels. Same as the Chair-CAD dataset, we obtained the transient latent vectors for all the golf scenes and one static latent vector for the training set. 

\subsection{Experimental Settings}
We implement our framework in PyTorch \cite{paszke2017automatic}. Please see supplementary material for details on implementation and values of different hyper-parameters ($D_\mathsf{s}, D_\mathsf{t}, \alpha$, etc.).

\textbf{Network Architecture.} We choose DCGAN \cite{radford2015unsupervised} as the generator architecture for the Chair-CAD and Golf scene dataset, and conditional generator architecture from \cite{mirza2014conditional} for the Weizmann Human Action dataset for our experiments. For the RNN, we employ a one-layer gated recurrent unit network with 500 hidden units \cite{chung2014empirical}. 

\textbf{Choice of Loss Function for }$\ell_{\text{rec}}$ \textbf{and} $\ell_{\text{static}}$\textbf{.} One straight forward loss function that can be used is the mean squared loss, but it has been shown in literature that it leads to generation of blurry pixels \cite{zhao2015loss}. Moreover, it has been shown empirically that generative functions in adversarial learning focus on edges \cite{bojanowski2018optimizing}. Motivated by this, the loss function for $\ell_{\text{rec}}$ and $\ell_{\text{static}}$ is chosen to be the Laplacian pyramid loss $\mathcal{L}_{\text{Laplacian}}$ \cite{ling2006diffusion} defined as 
\begin{align*}
    \mathcal{L}_{\text{Laplacian}}\br{v, \hat{v}} = \sum_l2^{2^l}\vert\mathsf{L}^l\br{v}-\mathsf{L}^l\br{\hat{v}}\vert_1
\end{align*} 
where $\mathsf{L}^l\br{\cdot}$ is the \textit{l}-th level of the Laplacian pyramid representation of the input.

\textbf{Baselines.} We compare our proposed method with two adversarial methods. For Chair-CAD and Weizmann Human Action, we use MoCoGAN \cite{tulyakov2018mocogan} as the baseline, and for Golf scene dataset, we use VGAN \cite{vondrick2016generating} as the baseline. We use the publicly available code for MoCoGAN and VGAN, and set the hyper-parameters as recommended in the published work. We also compare two different versions of the proposed method by ablating the proposed loss functions. Note that, we couldn't compare our results with Video-VAE \cite{he2018probabilistic} using our performance measures (described below) as the implementation has not been made available by the authors, and to the best of our efforts we couldn't reproduce the results provided by them. 

\textbf{Performance measures.}
Past video generation works have been evaluated quantitatively on Inception score (IS) \cite{he2018probabilistic}. But, it has been shown that IS is not a good evaluation metric for pixel domain generation, as the maximal IS score can be obtained by synthesizing a video from every class or mode in the given data distribution \cite{lucic2018gans, barratt2018note, theis2016note}. Moreover, a high IS does not guarantee any confidence on the quality of generation, but only on the diversity of generation. Since a generative model trained using our proposed method can generate all videos using the learned latent dictionary\footnote{Direct video comparison seems straight forward for our approach as the corresponding one-to-one ground truth is known. However, for \cite{tulyakov2018mocogan, vondrick2016generating}, we do not know which video is being generated (action may be known e.g. \cite{tulyakov2018mocogan}) which makes such direct comparison infeasible and unfair.}, and for a fair comparison with baselines, we use the following two measures, similar to measures provided in \cite{tulyakov2018mocogan}. We also provide relevant bounds computed on real videos for reference. Note that arrows indicate whether higher $\br{\uparrow}$ or lower $\br{\downarrow}$ scores are better. 

\vspace{1mm}
\noindent (1) \textbf{Relative Motion Consistency Score (MCS $\downarrow$)}: Difference between consecutive frames captures the moving components, and hence motion in a video. So, firstly each frame in the generated video, as well as the ground-truth data, is represented as a feature vector computed using a VGG16 network \cite{simonyan2014very} pre-trained on ImageNet \cite{russakovsky2015imagenet} at the \texttt{relu3\_3} layer. Secondly, the averaged consecutive frame-feature difference vector for both set of videos is computed, denoted by $\hat{f}$ and $f$ respectively. Finally, the relative MCS is then given by $\log_{10}\br{\Vert f - \hat{f}\Vert_2^2}$. 

\vspace{1mm}
\noindent (2) \textbf{Frame Consistency Score (FCS $\uparrow$)}: 
This score measures the consistency of the static portion of the generated video frames. We keep the first frame of the generated video as reference and compute the averaged structural similarity measure for all frames. The FCS is then given by the average of this measure over all videos.    

\subsection{Qualitative Results} 
\begin{figure*}[t!]
    \centering
    \begin{subfigure}[t]{0.31\textwidth}
      \hspace{-2em}
        \includegraphics[width=1.1\textwidth, height=1.5in]{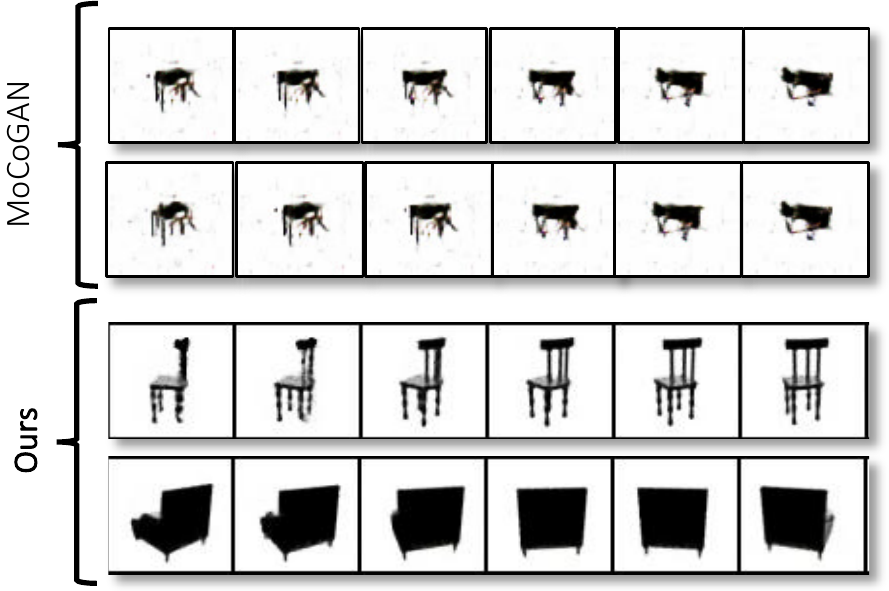}
        \caption{Chair-CAD \cite{aubry2014seeing}}
    \end{subfigure}%
    ~
    \begin{subfigure}[t]{0.31\textwidth}
        \hspace{-0.7em}
        \includegraphics[width= 1.1\textwidth, height=1.5in]{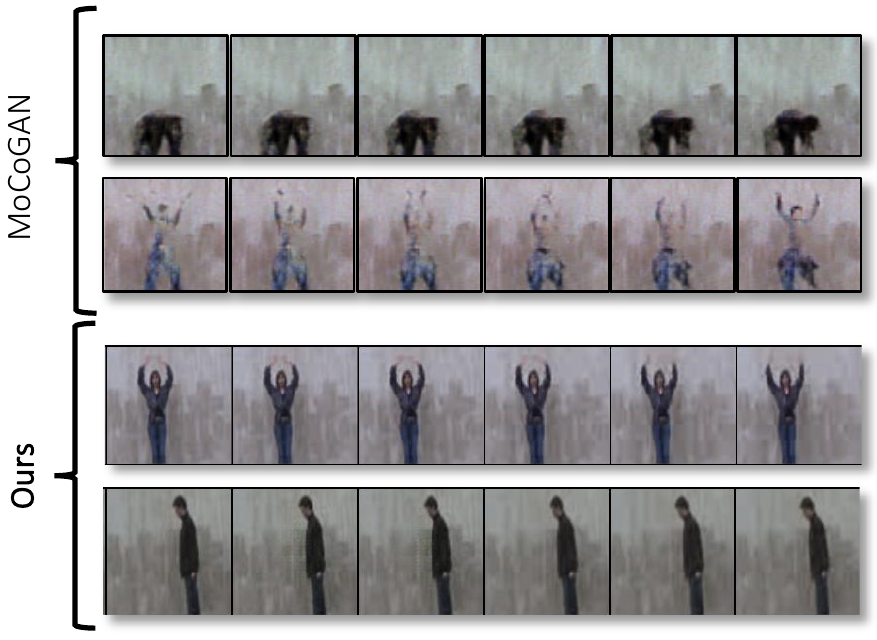}
        \caption{Weizmann Human Action \cite{gorelick2007actions}}
    \end{subfigure}
    ~
    \begin{subfigure}[t]{0.31\textwidth}
        \hspace{0.3em}
        \includegraphics[width= 1.1\textwidth, height=1.5in]{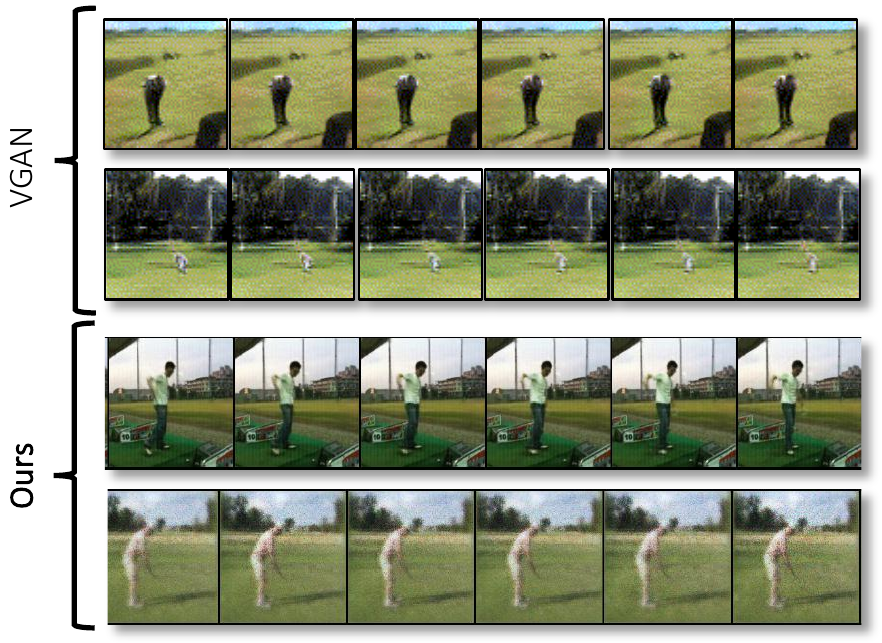}
        \caption{Golf \cite{vondrick2016generating}}
    \end{subfigure}
    \caption{\textbf{Qualitative results comparison with state-of-the-art methods.} We show two generated video sequences for MoCoGAN \cite{tulyakov2018mocogan} (for (a) Chair-CAD \cite{aubry2014seeing}, (b) Weizmann Human action  \cite{gorelick2007actions}), VGAN \cite{vondrick2016generating} (for (c) Golf scene \cite{vondrick2016generating}) (\textit{top}), and the proposed method (\textbf{Ours}, \textit{bottom}). The proposed method produces visually sharper, and consistently better using the non-adversarial training protocol. More examples have been provided in the supplementary material.}
    \label{fig:qual_result}
    \vspace{-1em}
\end{figure*}

Fig.~\ref{fig:qual_result} shows some examples with randomly selected frames of generated videos for the proposed method and the adversarial approaches MoCoGAN \cite{tulyakov2018mocogan} and VGAN \cite{vondrick2016generating}. For Chair-CAD \cite{aubry2014seeing} and Weizmann Human Action \cite{gorelick2007actions} dataset, it can be seen that the proposed method is able to generate visually good quality videos with a non-adversarial training protocol, whereas MoCoGAN produces blurry and inconsistent frames. Since we use optimized latent vectors unlike MoCoGAN (which uses random latent vectors for video generation), our method produces visually more appealing videos. Fig.~\ref{fig:qual_result} presents two particularly important points. As visualized for the Chair-CAD videos, the adversarial approach of MoCoGAN produces not only blurred chair images in the generated video, but they are also non-uniform in quality. Further, it can be seen that as the time step increases, MoCoGAN tends to generate the same chair for different videos. This shows a major drawback  of the adversarial approaches, where they fail to learn the diversity of the data distribution. Our approach overcomes this by producing a optimized dictionary of latent vectors which can be used for generating any video in the data distribution easily. To further validate our method for qualitative results, we present the following experiments.
\vspace{-0.25cm}
\subsubsection{Qualitative Ablation Study}
\begin{figure*}
    \centering
    \includegraphics[width=0.8\textwidth, height=1in]{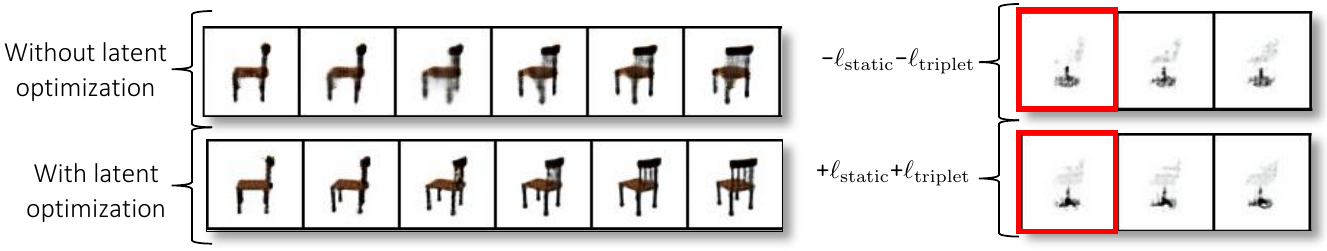}
    \caption{\textbf{Qualitative ablation study on Chair-CAD \cite{aubry2014seeing}.} \textit{Left}: It can be seen that the model is not able to generate good quality frames properly resulting in poor videos when the input latent space is not optimized, whereas with latent optimization, the generated frames are sharper. \textit{Right}: The impact of $\ell_{\text{static}}$ and $\ell_{\text{triplet}}$ is indicated by the red bounding boxes. Our method with $\ell_{\text{static}}$ and $\ell_{\text{triplet}}$ captures the difference between the white background and white chair, whereas without these two loss functions, the chair images are not distinguishable from their background. \texttt{+} and \texttt{-} indicate presence and absence of the terms, respectively.}
    \label{fig:qual_abl_result}
\end{figure*}

 Fig.~\ref{fig:qual_abl_result} qualitatively shows the contribution of the specific parts of the proposed method on Chair-CAD \cite{aubry2014seeing}. First, we investigate the impact of input latent vector optimization. For a fair comparison, we optimize the model for same number of epochs. It can be observed that the model benefits from the joint optimization of input latent space to produce better visual results. Next, we validate the contribution of $\ell_{\text{static}}$ and $\ell_{\text{triplet}}$ on a difficult video example whose chair color matches with the background. Our method, combined with $\ell_{\text{static}}$ and $\ell_{\text{triplet}}$, is able to distinguish between the white background and the white body of the chair model. 
\begin{table*}[ht]
\centering
\begin{subtable}{0.3\textwidth}
\centering \small{
\begin{tabular}{c|c|c}
\toprule[1.2pt]
    & MCS $\downarrow$  & FCS $\uparrow$ \\ \midrule
    Bound &  0.0 & 0.91\\
    \hline
MoCoGAN \cite{tulyakov2018mocogan} &   4.11   &  0.85 \\ \hline
Ours ($\texttt{-} \ell_{\text{triplet}} \texttt{-} \ell_{\text{static}}$)  &  3.83   &  0.77 \\ \hline
Ours ($\texttt{+} \ell_{\text{triplet}} \texttt{+} \ell_{\text{static}}$)  &   \textbf{3.32}  &  \textbf{0.89} \\ \bottomrule[1.2pt]
\end{tabular}}
\caption{Chair-CAD \cite{aubry2014seeing}}
\label{tab:quant_chair_cad}
\end{subtable}%
\qquad
\begin{subtable}{0.3\textwidth}
\centering \small{
\begin{tabular}{c|c|c}
\toprule[1.2pt]
    & MCS $\downarrow$  & FCS $\uparrow$ \\ \midrule
    Bound &  0.0 & 0.95\\
    \hline
MoCoGAN \cite{tulyakov2018mocogan} &   3.41   &  0.85 \\ \hline
Ours ($\texttt{-} \ell_{\text{triplet}} \texttt{-} \ell_{\text{static}}$)  &  3.87   &  0.79 \\ \hline
Ours ($\texttt{+} \ell_{\text{triplet}} \texttt{+} \ell_{\text{static}}$)  &   \textbf{2.63}  &   \textbf{0.90}\\ \bottomrule[1.2pt]
\end{tabular}}
\caption{Weizmann Human Action \cite{gorelick2007actions}} 
\label{tab:quant_weizmann}
\end{subtable}
\qquad
\begin{subtable}{0.3\textwidth}
\centering \small{
\begin{tabular}{c|c|c}
\toprule[1.2pt]
    & MCS $\downarrow$  & FCS $\uparrow$ \\ \midrule
    Bound &  0.0 & 0.97 \\
    \hline
VGAN \cite{vondrick2016generating} &  3.61    &  \textbf{0.88} \\ \hline
Ours ($\texttt{-} \ell_{\text{triplet}} \texttt{-} \ell_{\text{static}}$)  &  3.78    &  0.84 \\ \hline
Ours ($\texttt{+} \ell_{\text{triplet}} \texttt{+} \ell_{\text{static}}$)  &   \textbf{2.71}   &    0.84\\ 
\bottomrule[1.2pt]
\end{tabular}}
\caption{Golf \cite{vondrick2016generating}} 
\label{tab:golf}
\end{subtable}
\caption{\textbf{Quantitative results comparison with state-of-the-art methods.} We obtained better scores on the proposed method on both Chair-CAD \cite{aubry2014seeing}, Weizmann Human Action \cite{gorelick2007actions}, and Golf \cite{vondrick2016generating} datasets, compared to the adversarial approaches (MoCoGAN, and VGAN). Best scores have been highlighted in bold. }
\label{tab:main-quant}
\vspace{-1em}
\end{table*}

\begin{table}[H]
\centering 
\vspace{-1em}
\small{
\begin{tabular}{c|c|c|c|c|c|c|c|c|c}
\toprule[1.2pt]
\multicolumn{1}{c|}{\multirow{2}{*}{Actions}} & \multicolumn{9}{c}{Identities}\\
\cline{2-10}
     & P1  & P2 & P3  & P4 & P5  & P6  &  P7  &  P8  & P9 \\ \midrule
run  &$\bullet$   &   $\bullet$  &  $\bullet$    &    &   \textcolor{red}{ $\bullet$}   &     &      &      &    \\ \hline
walk &  \textcolor{green(ryb)}{ $\bullet$}   &  $\bullet$   &     &    &   $\bullet$   &     &    $\bullet$   &      &   \\ \hline
jump &     &   $\bullet$  &     &   $\bullet$  &   $\bullet$   &  $\bullet$    &   \textcolor{blue}{ $\bullet$}    &  $\bullet$     &   $\bullet$  \\ \hline
skip &   $\bullet$   &   $\bullet$  &    $\bullet$  &    &     &  $\bullet$    &   $\bullet$    &    \textcolor{yellow}{ $\bullet$}   &   $\bullet$ \\ 
\bottomrule[1.2pt]
\end{tabular}}
\caption{\textbf{Generating videos by exchanging unseen actions by identities.} Each cell in this table indicates a video in the dataset. Only cells containing the symbol \textcolor{black}{$\bullet$} indicate that the video was part of the training set. We randomly generated videos corresponding to rest of the cells indicated by symbols \textcolor{red}{$\bullet$}, \textcolor{green(ryb)}{$\bullet$}, \textcolor{yellow}{$\bullet$}, and \textcolor{blue}{$\bullet$}, visualized in Fig.~\ref{fig:motion_exchange}.}
\label{tab:exchange}
\vspace{-0.7em}
\end{table}
\subsubsection{Action Exchange}
Our non-adversarial approach can effectively separate the static and transient portion of a video, and generate videos unseen during the training protocol. To validate these points, we choose a simple \textit{matrix} completion for the combination of identities and actions in the Weizmann Human action \cite{gorelick2007actions} dataset. For training our model, we created a set of videos (without any cropping to present the complete scale of the frame) represented by the cells marked with $\bullet$ in Tab.~\ref{tab:exchange}. 
\begin{figure}
    \centering
    \includegraphics[width=0.47\textwidth]{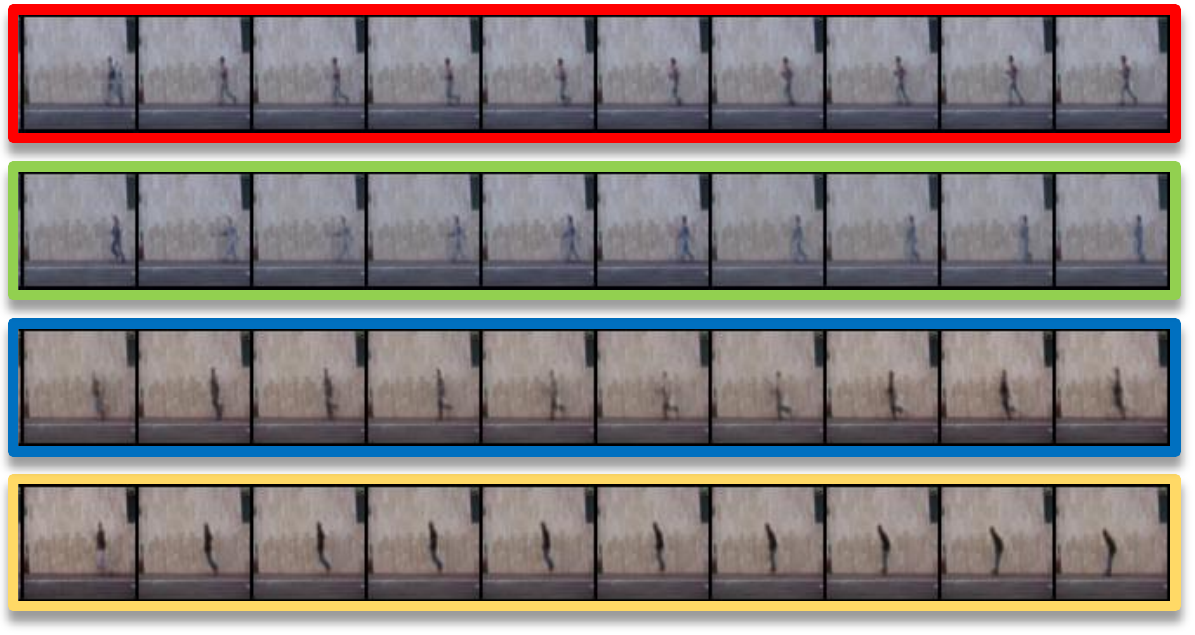}
    \caption{\textbf{Examples of action exchange to generate unseen videos.} This figure shows the generated videos unseen during the training of the model with colored bounding boxes indicating the colored dots (\textcolor{red}{ $\bullet$}, \textcolor{yellow}{ $\bullet$}, \textcolor{blue}{ $\bullet$}, \textcolor{green(ryb)}{ $\bullet$}) referred to in Tab.~\ref{tab:exchange}. This demonstrates the effectiveness of our method in disentangling static and transient portion of videos.}
    \label{fig:motion_exchange}
    \vspace{-1.25em}
\end{figure}
Hence, the unseen videos correspond to the cells not marked with $\bullet$. During testing, we randomly generated these unseen videos (marked with \textcolor{red}{ $\bullet$}, \textcolor{yellow}{ $\bullet$}, \textcolor{blue}{ $\bullet$} and \textcolor{green(ryb)}{ $\bullet$} in Tab.~\ref{tab:exchange}), and the visual results are shown in Fig.~\ref{fig:motion_exchange}. This experiment clearly validates our claim of static (identities) and transient (action) portion disentanglement of a video and, generation of unseen videos by using combinations of action and identities \textit{not} part of training set. Note that generated videos may not exactly resemble ground truth videos of the said combinations as we learn $\textbf{z}^{(\mathsf{t})}$ over a class of many videos. 
\vspace{-0.4cm}
\subsubsection{Frame Interpolation}\label{sec:Frame_inter}
\begin{figure}
    \centering
    \includegraphics[width=0.4\textwidth, height = 1.in]{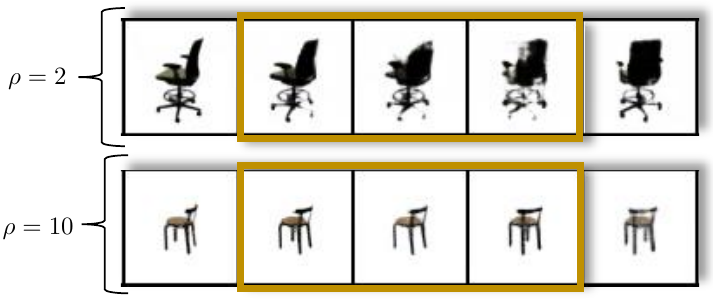}
    \caption{\textbf{Examples of frame interpolation.} An important advantage of our method is translation of interpolation in the learned latent space to video space using (\ref{7}). It can be observed that as $\rho$ increases, the interpolation (bounded by color) is better. Note that the adjacent frames are also generated frames, and not ground truth frames.}
    \label{fig:interpolation}
    \vspace{-0.5cm}
\end{figure}

To show our methodology can be employed for frame interpolation, we trained our model using the loss (\ref{7}) for $\rho = 2$ and $\rho = 10$. During testing, we generated intermediate frames by interpolating learned latent variables of two distinct frames. For this, we computed the difference $\Delta z^{(\mathsf{t})}$ between the learned latent vectors of second ($z^{(\mathsf{t})}_2$) and fifth ($z^{(\mathsf{t})}_5$) frame, and generated $k=3$ unseen frames using $\{z_2^{(\mathsf{t})} + \sfrac{n}{k}\Delta z^{(\mathsf{t})} \}_{n=1}^k$, after concatenating with $z^{(\mathsf{s})}$. Fig.~\ref{fig:interpolation} shows the results of interpolation between second and fifth frames for two randomly chosen videos. Thus, our method is able to produce dynamically consistent frames with respect to the reference frames without any pixel clues.

\subsection{Quantitative Results}
Quantitative result comparisons with respect to baselines have been provided in Tab.~\ref{tab:main-quant}. Compared to videos generated by the adversarial method MoCoGAN \cite{tulyakov2018mocogan}, we report a relative decrease of 19.22\% in terms of MCS, and 4.70\% relative increase in terms of FCS for chair-CAD dataset \cite{aubry2014seeing}. For the Weizmann Human Action \cite{gorelick2007actions} dataset, the proposed method is reported to have a relative decrease of 22.87\% in terms of in terms of MCS, and 4.61\% relative increase in terms of FCS. Similarly for Golf scene dataset \cite{vondrick2016generating}, we perform competitively with VGAN \cite{vondrick2016generating} with a observed relative decrease of 24.90\% in terms of in terms of MCS. A important conclusion from these results is that our proposed method, being non-adversarial in nature, learns to synthesize a diverse set of videos, and is able to perform at par with adversarial approaches. It should be noted that a better loss function for $\ell_{\text{rec}}$ and $\ell_{\text{static}}$ would produce stronger results. We leave this for future works.  
\vspace{-1em}
\subsubsection{Quantitative Ablation Study}
\begin{table}
\small{
\begin{subtable}[t]{0.45\hsize}
\centering
\begin{tabular}{c|c|c}
\toprule[1.2pt]
 &  MCS $\downarrow$ & FCS $\uparrow$  \\
\midrule
Bound &   0   & 0.91  \\ \hline
$\texttt{-}\mathsf{Z}$ & 3.96 & 0.75   \\
\hline
 $\texttt{+}\mathsf{Z}$ & 3.32 & 0.89   \\
\bottomrule[1.2pt]
\end{tabular}
\caption{With respect to latent space optimization.}
\label{tab:ablation_z}
\end{subtable}}~
\small{
\begin{subtable}[t]{0.5\hsize}
\centering
\begin{tabular}{c|c|c}
\toprule[1.2pt]
    & MCS $\downarrow$  & FCS $\uparrow$ \\ \midrule
Bound               &  0   &  0.91 \\ \hline
$\texttt{-} \ell_{\text{triplet}} \texttt{-} \ell_{\text{static}}$  &   3.83   & 0.77  \\ \hline
$\texttt{-} \ell_{\text{triplet}} \texttt{+} \ell_{\text{static}}$  &  3.82 & 0.85  \\ \hline
$\texttt{+} \ell_{\text{triplet}} \texttt{-} \ell_{\text{static}}$  &  3.36 &  0.81 \\ \hline
$\texttt{+} \ell_{\text{triplet}} \texttt{+} \ell_{\text{static}}$  &  3.32 &   0.89\\ 
\bottomrule[1.2pt]
\end{tabular}
\caption{With respect to loss functions}
\label{tab:ablation_loss}
\end{subtable}}
\caption{\textbf{Ablation study of proposed method on Chair-CAD} \cite{aubry2014seeing}. In (a), we evaluate contributions of latent space optimization ($\mathsf{Z}$). In (b), we evaluate contributions of $\ell_{\text{triplet}}$ and $\ell_{\text{static}}$ in four combinations. \texttt{+} and \texttt{-} indicate presence and absence of the terms, respectively.}
\vspace{-1em}
\end{table}
In this section, we demonstrate the contribution of different components in our proposed methodology on the Chair-CAD \cite{aubry2014seeing} dataset. For all the experiments, we randomly generate 500 videos using our model by using the learned latent vector dictionary. We divide the ablation study into two parts. Firstly, we present the results for impact of the learned latent vectors on the network modules. For this, we simply generate videos once with the learned latent vectors ($\texttt{+}\mathsf{Z}$), and once with randomly sampled latent vectors from a different distribution ($\texttt{-}\mathsf{Z}$). The inter-dependency of our model weights and the learned latent vectors can be interpreted from Tab.~\ref{tab:ablation_z}. We see that there is a relative decrease of 16.16\% in MCS from 3.96 to 3.32, and 18.66\% of relative increase in FCS. This shows that optimization of the latent space in the proposed method is important for good quality video generation.

Secondly, we investigate the impact of the proposed losses on the proposed method. Specifically, we look into four possible combinations of $\ell_{\text{triplet}}$ and $\ell_{\text{static}}$. The results are presented in Tab.~\ref{tab:ablation_loss}. It can observed that the combination of triplet loss $\ell_{\text{triplet}}$ and static loss $\ell_{\text{static}}$ provides the best result when employed together, indicated by the relative decrease of 14.26\% in MCS from 3.83 to 3.32.

\section{Conclusion}
We present a non-adversarial approach for synthesizing videos by jointly optimizing both network weights and input latent space.
Specifically, our model consists of a global static latent variable for content features, a frame specific transient latent variable, a deep convolutional generator, and a recurrent neural network which are trained using a regression-based reconstruction loss, including a triplet based loss.
Our approach allows us to generate a diverse set of almost uniform quality videos, perform frame interpolation, and generate videos unseen during training. Experiments on three standard datasets show the efficacy of our proposed approach over state-of-the-methods.

\textbf{Acknowledgements.} The work was partially supported by NSF grant 1664172 and ONR grant N00014-19-1-2264.
\FloatBarrier
{\small
\bibliographystyle{ieee_fullname}
\bibliography{egbib}
}

\onecolumn
\setcounter{section}{-5}
\begin{center}
  \Large\bf{Non-Adversarial Video Synthesis with Learned Priors\\ (Supplementary Material)}  
\end{center}

\date{\vspace{-3ex}}

\renewcommand\thesection{\Alph{section}}
\setcounter{section}{0}
\begin{table} [h]
\begin{center}
\begin{tabular}{ c|l }
\toprule[1.2pt]
 \textbf{Page} \#  & \textbf{Content} \\
 \hline
 \hline
 \texttt{12}  & \textbf{Dataset Descriptions} \\
 \hline
 \texttt{12}  & \textbf{Implementation Details} 
     \\
     &\qquad$\bullet$ Hyper-parameters  \\
     &\qquad$\bullet$ Other details  \\
 \hline
 \texttt{13}  & \textbf{More Qualitative Examples} \\
     &\qquad$\bullet$ Qualitative examples on Chair-CAD \cite{aubry2014seeing}  \\
     &\qquad$\bullet$ Qualitative examples on Weizmann Human Action \cite{gorelick2007actions}  \\
     &\qquad$\bullet$ Qualitative examples on Golf scene \cite{vondrick2016generating}  \\
     &\qquad$\bullet$ Interpolation examples on Chair-CAD \cite{aubry2014seeing} and Weizmann Human Action \cite{gorelick2007actions} \\
 \bottomrule[1.2pt]
\end{tabular}
\end{center}
\caption{Supplementary Material Overview.}
\end{table}
\newpage
\section{Dataset Descriptions}
\begin{figure}[ht]
    \centering
    \includegraphics[scale=0.5]{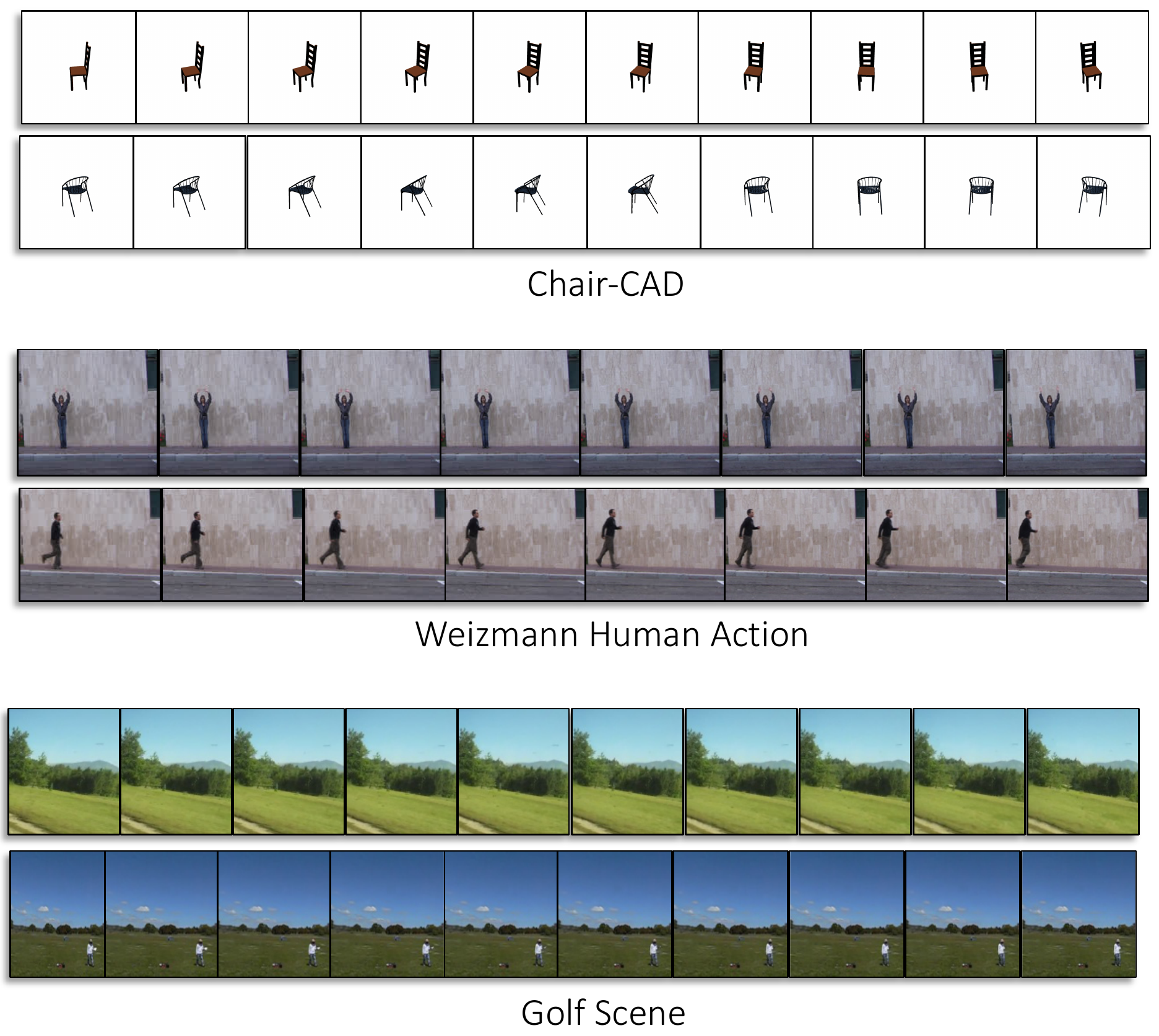}
    \caption{\textbf{Sample videos from datasets used in the paper.} Two unprocessed video examples from Chair-CAD \cite{aubry2014seeing}, Weizmann Human Action \cite{mirza2014conditional}, and Golf Scene \cite{vondrick2016generating} datasets have been presented here. As seen from the examples, datasets are diverse in nature, different in categories and present unique challenges in learning the transient and static portions of the videos. Best viewed in color.}
    \label{fig:supp_raw_data}
    \vspace{-1em}
\end{figure}

\textbf{Chair-CAD} \cite{aubry2014seeing}\textbf{.} This dataset provides 1393 chair-CAD models. Each model frame sequence is produced using two elevation angles in addition to thirty one azimuth angles. All the chair models have been designed to be at a fixed distance with respect to the camera. The authors provide four video sequences per CAD model. We choose the first 16 frames of each video for our paper, and consider the complete dataset as one class.

\textbf{Weizmann Human Action} \cite{gorelick2007actions}\textbf{.} This dataset is a collection of 90 video sequences showing nine different identities performing 10 different actions, namely, \texttt{run}, \texttt{walk}, \texttt{skip}, \texttt{jumping-jack} (or `jack'),  \texttt{jump-forward-on-two-legs} (or `jump'), \texttt{jump-in-place-on-two-legs} (or `pjump'), \texttt{gallopsideways} (or `side'), \texttt{wave-two-hands} (or `wave2'), \texttt{waveone-hand} (or `wave1'), and \texttt{bend}. We randomly choose 16 consecutive frames for every video in each iteration during training.

\textbf{Golf Scene} \cite{vondrick2016generating}.
\cite{vondrick2016generating} released a dataset containing 35 million clips (32 frames each) stabilized by SIFT+RANSAC. It contains several categories filtered by a pre-trained Place-CNN model, one of them being the Golf scenes. The Golf scene dataset contains 20,268 golf videos. Due to many non-golf videos being part of the golf category (due to inaccurate labels), this dataset presents a particularly challenging data distribution for our proposed method. Note that for a fair comparison, we further selected our training set videos from this provided dataset pertaining to golf action as close as possible. We then trained the VGAN \cite{vondrick2016generating} model on this selected videos for a fair comparison.

\section{Implementation Details}
We used Pytorch \cite{paszke2017automatic} for our implementation. The Adam optimizer \cite{kingma2014adam}, with $\epsilon = 10^{-8}$, $\beta_1 = 0.9$ and $\beta_2 = 0.999$, was used to update the model weights and SGD optimizer \cite{ruder2016overview}, with momentum $ = 0.9$, was used to update the latent spaces. The corresponding learning rate for the generator $\tau_{\scaleto{\mathcal{G}}{4pt}}$, the RNN $\tau_{\scaleto{\mathcal{R}}{4pt}}$, and the latent spaces $\tau_{\scaleto{\mathcal{Z}_{\mathsf{V}}}{4pt}}$ were set as values indicated in Tab.~\ref{tab:hyper}.

\textbf{Hyper-parameters.} \cite{tulyakov2018mocogan, vondrick2016generating, saito2017temporal} that generate videos from latent priors have no dataset split as the task is to synthesize high quality videos from the data distribution, and then evaluate the model performance. All hyperparameters (except $D_{\mathsf{s}}$, $D_{\mathsf{t}}$) are set as described in \cite{sermanet2018time, schroff2015facenet, tulyakov2018mocogan, wang2018video} (e.g. $\alpha$ from \cite{sermanet2018time}). For $D_{\mathsf{s}}$ and $D_{\mathsf{t}}$, we follow the strategy used in Sec. 4.3 of \cite{tulyakov2018mocogan} and observe that our model generates videos with good visual quality (FCS) and plausible motion (MCS) for Chair-CAD when ($D_{\mathsf{s}}$, $D_{\mathsf{t}}$) = (206, 50). Same strategy is used for all datasets. The hyper-parameters employed with respect to each dataset used in this paper is given in Tab.~\ref{tab:hyper}. $\mathcal{G}$ and $\mathcal{R}$ refer to the generator, with weights $\gamma$, and RNN, with weight $\theta$, respectively. $\tau_{(\cdot)}$ represents the learning rate. $\mu_{(\cdot)}$ represents the number of epochs. $D_\mathsf{s}$ and $D_\mathsf{t}$ refer to the static and transient latent dimensions, respectively. $\lambda_{\text{s}}$ and $\lambda_{\text{t}}$ refer to the static loss, and triplet loss regularization constants, respectively. $\alpha$ is the margin for triplet loss. $l$ refers to the level of the Laplacian pyramid representation used in $\ell_{\text{rec}}$ and $\ell_{\text{static}}$.
\begin{table}[ht]
    \centering
    \small{
    \begin{tabular}{l|c|c|c|c|c|c|c|c|c|c|c}
\toprule[1.2pt]
\multicolumn{1}{c|}{\multirow{2}{*}{Datasets}} & \multicolumn{9}{c}{Hyper-parameters}\\
\cline{2-12}
     & $D_\mathsf{s}$  & $D_\mathsf{t}$ & $\lambda_{\text{s}}$  & $\lambda_{\text{t}}$ & $\alpha$  &  $\tau_{\scaleto{\mathcal{G}}{4pt}}$  &  $\tau_{\scaleto{\mathcal{R}}{4pt}}$  & $\tau_{\scaleto{\mathcal{Z}_{\mathsf{V}}}{4pt}}$ & $\mu_\gamma$ & $\mu_{\scaleto{\mathcal{Z}_{\mathsf{V}}, \theta}{4pt}}$ & $l$\\ \midrule
Chair-CAD \cite{aubry2014seeing}  
& 206   &  50   &  0.01  &  0.01  &  2  & $6.25 \times 10^{-5}$ & $6.25 \times 10^{-5}$ & 12.5   &  5  & 300 & 4  \\ \hline
Weizmann Human Action \cite{gorelick2007actions}
& 56    &  200  &  0.01  &  0.1   &  2  & $6.25 \times 10^{-5}$ & $6.25 \times 10^{-3}$ &  12.5  &  5   &  700  & 3  \\ \hline
Golf Scene \cite{vondrick2016generating}
&  56   &  200  &  0.01  &  0.01  &  2  &         0.1          &  0.1 &  12.5      & 10 &  1000  & 4 \\
\bottomrule[1.2pt]
\end{tabular}}
    \caption{Hyper-parameters used in all experiments for all datasets.}
    \label{tab:hyper}
\end{table}

\textbf{Other details.} We performed all our experiments on a system with 48 core Intel(R) Xeon(R) Gold 6126 processor with 256GB RAM. We used NVIDIA GeForce RTX 2080 Ti for all GPU computations during training. Further, NVIDIA Tesla K40 GPUs were used for computation of all evaluation metrics in our experiments. All our implementations are based on non-optimized PyTorch based codes. Our runtime analysis revealed that it took on average one to two days to train the model and obtain learned latent vectors.


\section{More Qualitative Examples}
\begin{figure}[H]
    \centering
    \includegraphics[scale=0.7]{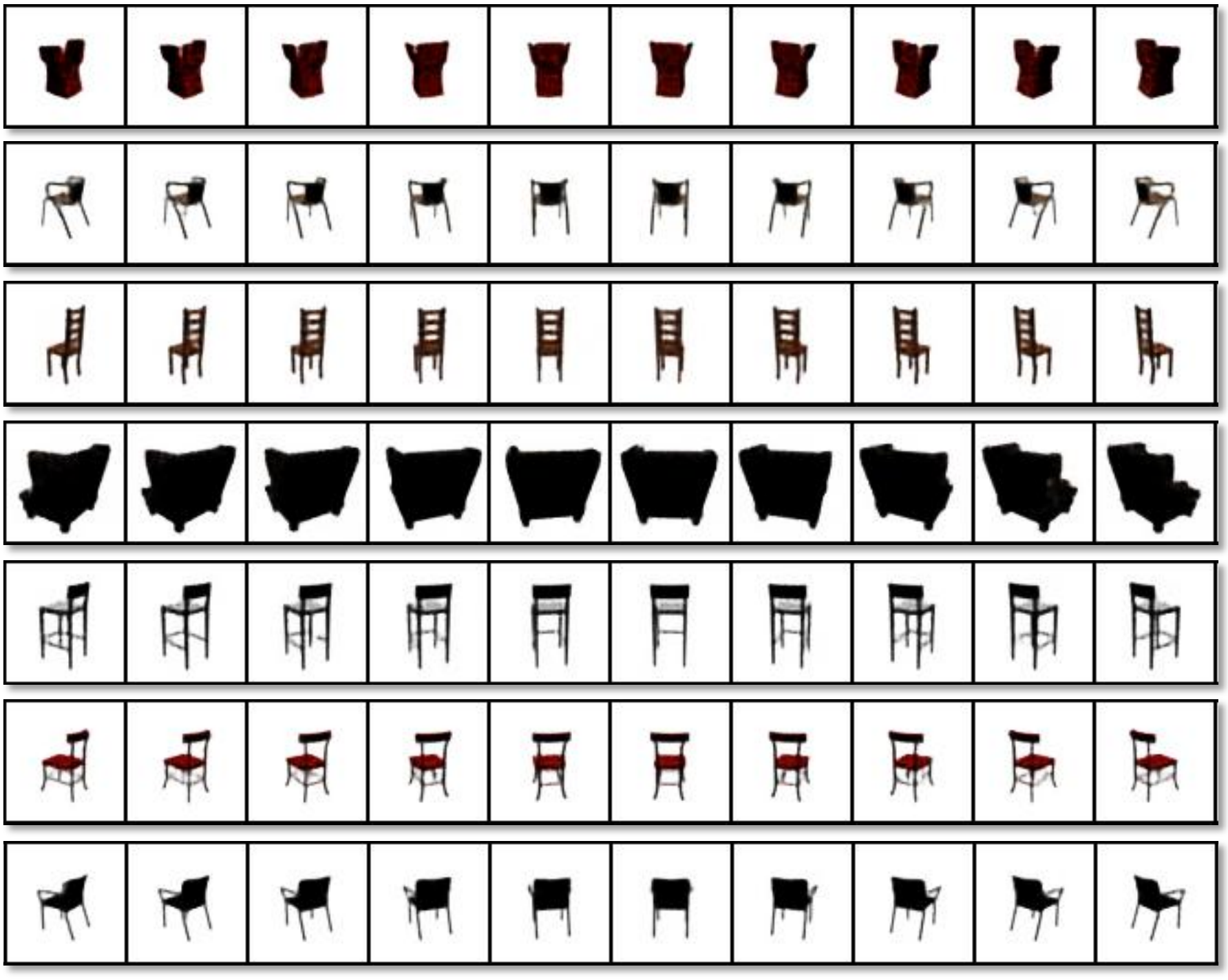}
    \vspace{-0.8em}
    \caption{\textbf{Qualitative results on Chair-CAD}  \cite{aubry2014seeing}. On this large scale dataset, our model is able to capture the intrinsic rotation and color of videos unique to each chair model. This shows the efficacy of our approach, compared to adversarial approaches such as MoCoGAN \cite{tulyakov2018mocogan} which produce the same chair for all videos, with blurry frames (See Fig.~1 of main manuscript).}
    \label{fig:supp_chair_cad}
    \vspace{-0.4em}
\end{figure}

\begin{figure}[H]
    \vspace{-1.9em}
    \centering
    \includegraphics[scale=0.7]{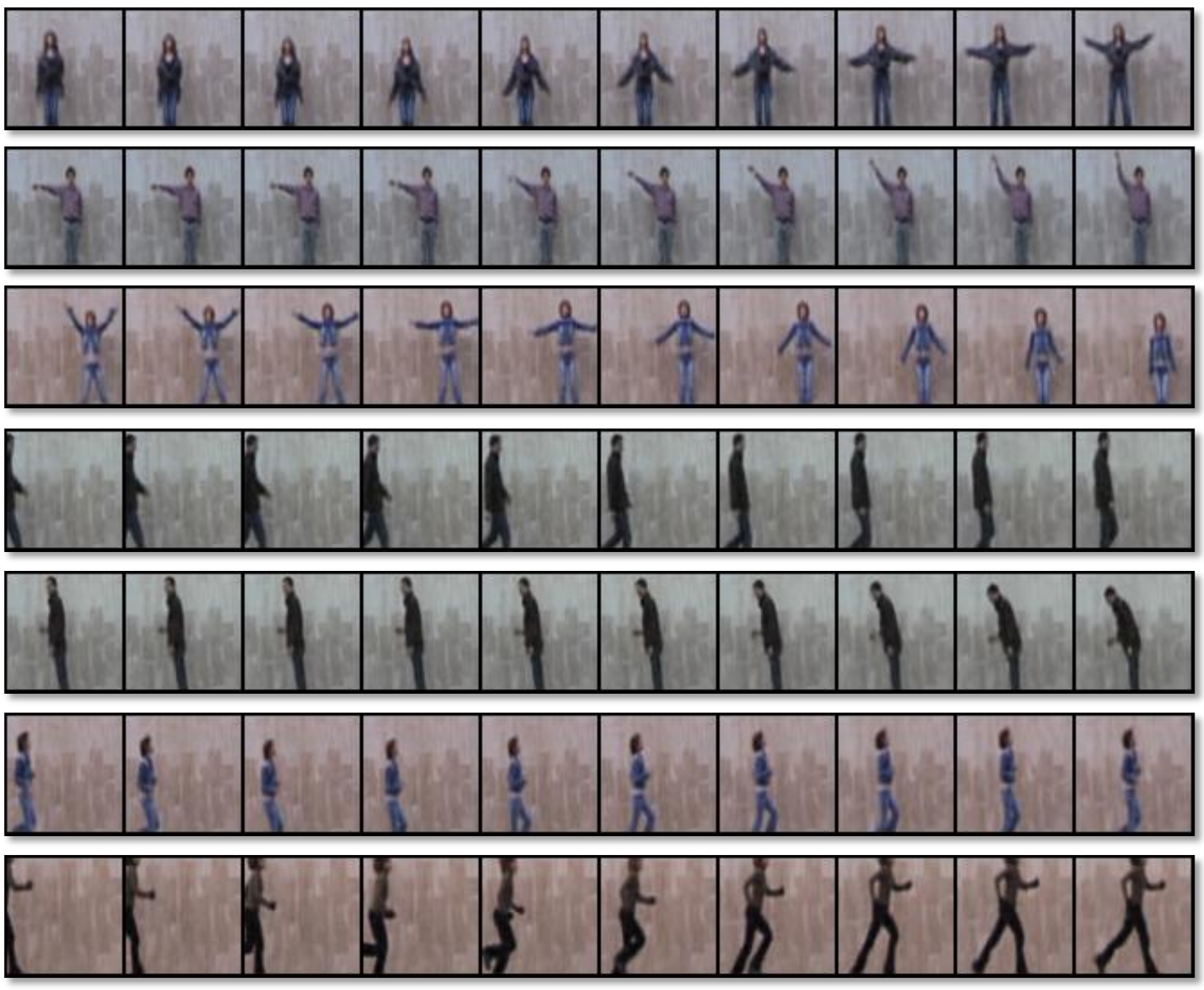}
    \vspace{-0.8em}
    \caption{\textbf{Qualitative results on Weizmann Human Action} \cite{gorelick2007actions}. The videos show that our model produces sharp visual results with the combination of trained generator, RNN along with 9 identities, and 10 different action latent vectors.}
    \label{fig:supp_wz}
\end{figure}

\begin{figure}[H]
\vspace{-1.5em}
    \centering
    \includegraphics[scale=0.7]{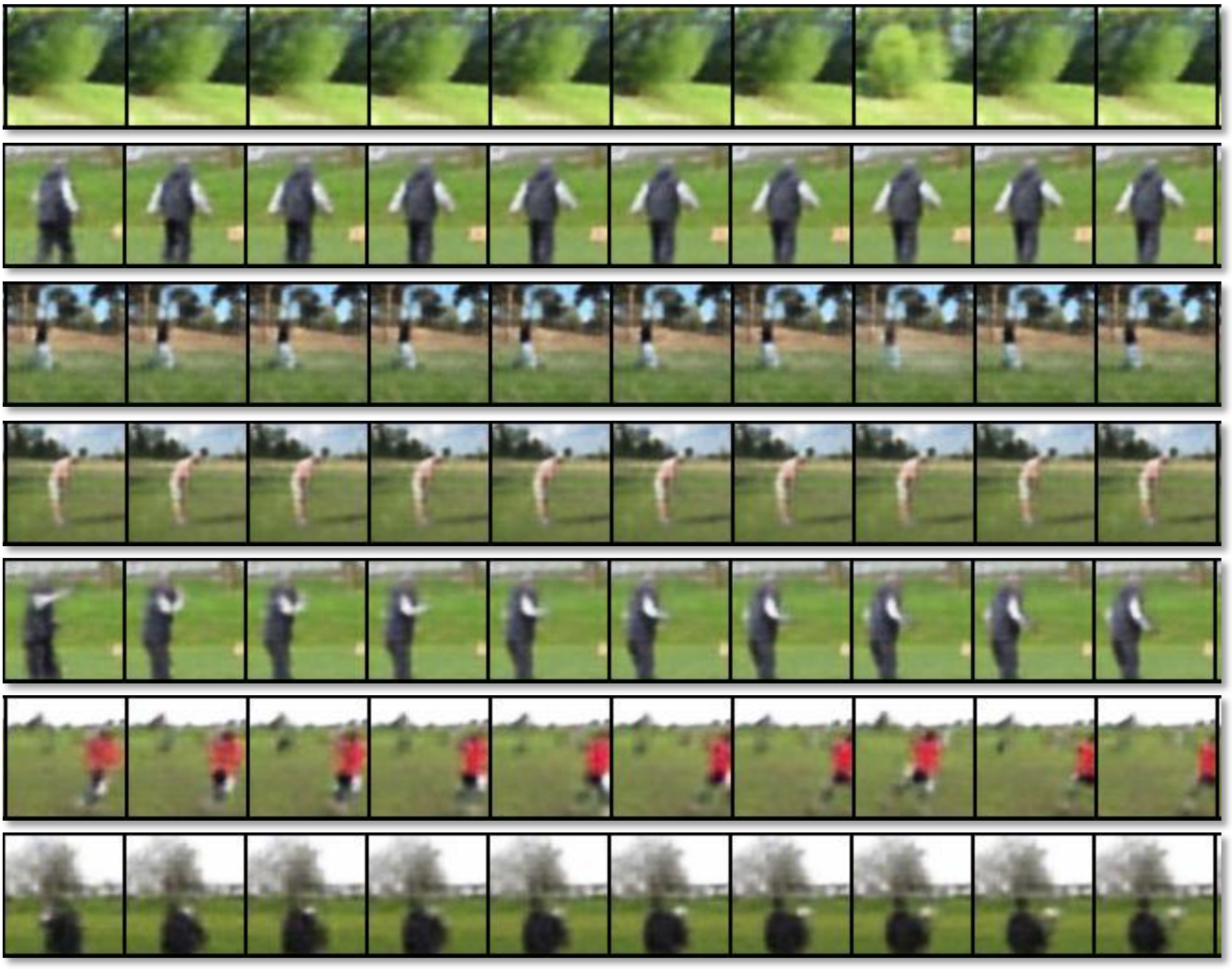}
    \vspace{-0.2em}
    \caption{\textbf{Qualitative results on Golf Scene}
    \cite{vondrick2016generating}. Our proposed approach produces visually good results on this particularly challenging dataset. Due to incorrect labels on the videos, this dataset has many non-golf videos. Our model is still able to capture the static and transient portion of the videos, although better filtering can still improve our results.}
    \label{fig:supp_golf}
\end{figure}

\begin{figure}[H]
    \begin{subfigure}[t]{0.45\textwidth}
         \centering
    \includegraphics[scale=1]{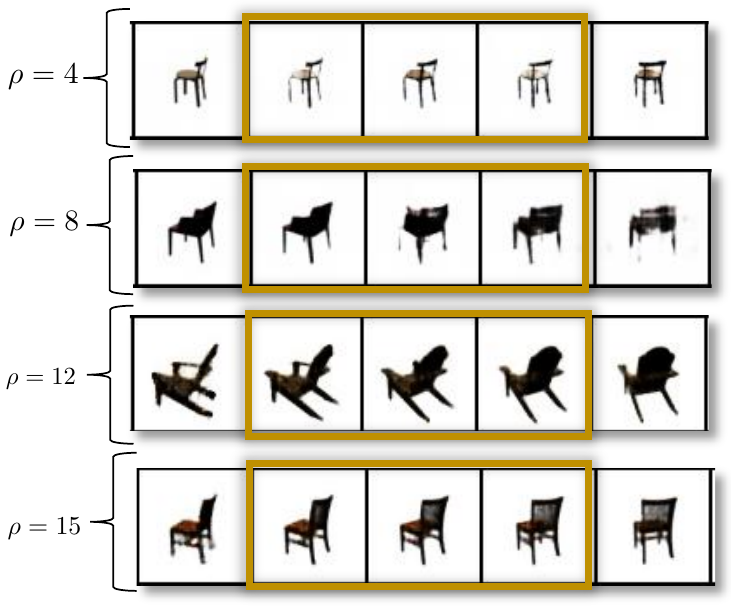}
    \vspace{-0.5em}
    \caption{Chair-CAD \cite{aubry2014seeing}}
    \label{fig:supp_inter_cc}
    \end{subfigure}
    \hfill
    \begin{subfigure}[t]{0.45\textwidth}
         \centering
    \includegraphics[scale=1]{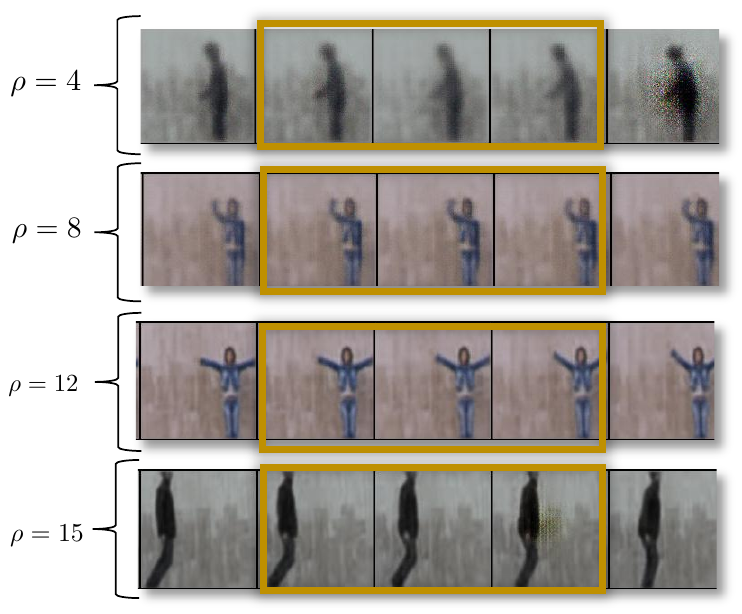}
    \vspace{-0.5em}
    \label{fig:supp_inter_wz}
    \caption{Weizmann Human Action \cite{gorelick2007actions}}
    \end{subfigure}
    \caption{\textbf{More interpolation results}. In this figure, $\rho$ represents the rank of transient latent vectors $\textbf{z}_\mathsf{t}$. We present the interpolation results on (a) Chair-CAD dataset, and (b) Weizmann Human Action, for different values of $\rho$. It can be observed that as $\rho$ increases, the interpolation becomes clearer.}
    \label{fig:supp_inter}
\end{figure}

In this section, we provide more qualitative results of generated videos synthesized using our proposed approach on each dataset (Fig.~\ref{fig:supp_chair_cad} for Chair-CAD \cite{aubry2014seeing} dataset, Fig.~\ref{fig:supp_wz} for Weizmann Human Action \cite{gorelick2007actions} dataset, and Fig.~\ref{fig:supp_golf} for Golf scene \cite{vondrick2016generating} dataset). We also provide more examples interpolation experiment in Fig.~\ref{fig:supp_inter}. 

\vspace{3mm}
\end{document}